\documentclass[trackchanges]{aastex701}
\usepackage{CJK}
\usepackage[T1]{fontenc}
\usepackage{enumitem} 
\usepackage{amsmath} 
\usepackage{amssymb} 
\usepackage{natbib}
\usepackage{bm}
\usepackage{booktabs} 
\usepackage{multirow} 
\usepackage{array}    
\usepackage{colortbl} 
\usepackage{threeparttable}
\usepackage{diagbox}
\usepackage{tikz}
\usepackage{siunitx}
\usepackage{makecell}

\newcommand{\Mbh}{M_{\bullet}}
\newcommand{\mbh}{m_{\bullet}}

\newcommand{\rbh}{r_{\rm h\bullet}}
\newcommand{\Ms}{M_{\star}}
\newcommand{\rs}{r_{\rm h\star}}
\newcommand{\mbhmax}{m_{\bullet,\rm max}}
\newcommand{\mbhmin}{m_{\bullet,\rm min}}
\newcommand{\vk}{v_{\mathrm{k}}}
\newcommand{\mbhf}{m_{\rm f\bullet}}
\newcommand{\mbhave}{\overline{\mbh} }
\newcommand{\mave}{\overline{m}}
\newcommand{\msave}{\overline{m_\star}}
\newcommand{\rc}{r_{\mathrm{c}}}
\newcommand{\tcoll}{t_{\mathrm{c}}}
\newcommand{\tGW}{t_{\mathrm{GW}}}
\newcommand{\rcbh}{r_{\rm c\bullet}}

\newcommand{\rhozstar}{\rho_{0.1\,\star}}

\begin{document}
\begin{CJK*}{UTF8}{gbsn}
\title{Rapid intermediate-mass black hole formation via runaway mergers of black holes}

\author[gname=Yining,sname=Sun]{Yining Sun}
\affiliation{School of Physics and Astronomy, Sun Yat-sen University, Daxue Road, Zhuhai, 519082, China}
\email{}

\author[orcid=0000-0001-8713-0366,sname='Wang']{Long Wang}
\affiliation{School of Physics and Astronomy, Sun Yat-sen University, Daxue Road, Zhuhai, 519082, China}
\affiliation{CSST Science Center for the Guangdong-Hong Kong-Macau Greater Bay Area, Zhuhai, 519082, China}
\email[show]{wanglong8@sysu.edu.cn}  





\begin{abstract}
Observations indicate that supermassive black holes (SMBHs) in high-redshift galaxies formed on timescales far shorter than classical growth models allow. 
One hypothesis suggests intermediate-mass black hole (IMBH) seeds as an efficient growth channel. 
Using $N$-body simulations, we demonstrate that in dense stellar-mass black hole (BH) clusters ($\ge 5\times10^9~M_{\odot}/{\rm pc}^3$), runaway gravitational-wave binary BH (BBH) mergers can produce a $\sim10^3~M_\odot$ IMBH within 10~Myr from the formation of the BH subsystem. This scenario is simple and avoids large uncertainties regarding stellar mergers and evolution in the IMBH formation via very massive stars channel.
We find that the runaway GW-merger mechanism relies on hard BBH formation through a chain of exchanged soft BBHs with accumulated hardening, which is far more efficient than three-body scattering.  We analyze how IMBH formation depends on cluster density, total mass, initial mass function, and stellar halo potential. We find that due to cluster expansion, the systems forming IMBHs have densities consistent with present-day nuclear star clusters, such as those in the Milky Way and M33. Furthermore, we show that IMBH spin remains low due to repeated mergers, and we estimate the rate of GW190521 and GW231123-like events within the first 100~Myr  to be $2.27-247.52$ and $3.23-63.63 $ per Gyr per cluster. 

\end{abstract}



\section{Introduction}

Observations of high-redshift quasars ($z>6$) indicate that supermassive black holes (SMBHs) exceeding $10^9 M_\odot$ already formed within the first billion years after the Big Bang. Over 200 quasar-like samples with $z>6$ were discovered, indicating that SMBHs are common at high redshifts \citep{2016ApJS..227...11B,2017ApJ...849...91M,2017MNRAS.466.4568T,2017NatSR...741617K,2018ApJ...854...97D,2019ApJ...874L..30V,2016ApJ...819...24W,2017ApJ...839...27W,2019ApJ...884...30W,2017MNRAS.468.4702R,2019MNRAS.487.1874R,2019ApJ...870L..11F,2019AJ....157..236Y,2015ApJ...798...28K,2018ApJS..237....5M,2018PASJ...70S..35M,2019ApJ...883..183M,2019ApJ...872L...2M}. Notable examples include one of the most massive known SMBHs, SDSS~J010013.02+280225.8, with a mass of $1.2 \times 10^{10} M_\odot$ at $z=6.3$ \citep{2015Natur.518..512W}, and a $10^7 \sim 10^8 M_{\odot}$ SMBH at $z=10.3$ \citep{2024NatAs...8..126B}. Other high-redshift SMBHs include ULAS J1342+0928 ($7.8 \times 10^8 M_{\odot}$) at $z=7.54$ \citep{2018Natur.553..473B} and a $(1.6 \pm 0.4) \times 10^9 M_\odot$ SMBH at $z = 7.642$ \citep{2021ApJ...907L...1W}. 
Following the discovery of ``little red dots'' (LRDs) \citep{2024ApJ...964...39G, 2024ApJ...963..129M}, several of these compact sources have also been identified as hosts of SMBHs. For instance, \citet{2023ApJ...957L...7K} found a $1.6 \times 10^9 M_\odot$ SMBH at $z=8.50$, \citet{2025ApJ...980L..29A} reported a $1.9 \times 10^7 M_\odot$ SMBH at $z=7.0371$.

%

To explain the rapid formation of SMBHs at high redshifts, several solutions have been proposed, generally classified into three pathways based on the seed mass \citep[see reviews in ][]{2020ARA&A..58...27I}:

\textit{The light seed pathway} ($10^{1-2} M_\odot$): A natural explanation is that these early SMBHs originated from the stellar-mass remnants (light seeds) of Population III (Pop III) stars.
If their growth were governed by Eddington-limited accretion with a radiative efficiency of approximately 10\% , which is consistent with typical lower-redshift SMBHs \citep[$z=2-3$; ][]{Haehnelt_1998,Yu_2002,2004MNRAS.354.1020S}, 
a 100 $M_\odot$ Pop III seed would need approximately 0.8 Gyr of uninterrupted accretion to reach $10^9 M_\odot$. However, several effects could prevent such continuous accretion, including feedback from the accreting BHs, ultraviolet radiation and supernovae from Pop III stars within the shallow gravitational potential wells of mini-halos \citep{2007MNRAS.374.1557J,2008ApJ...679..925W,Milosavljevi__2009,2009ApJ...701L.133A,2023MNRAS.518.3606S}. Therefore, 
efficiently building up massive SMBHs from light seeds requires super-Eddington accretion 
\citep[e.g.][]{1979MNRAS.187..237B,2004MNRAS.349...68B,2004ARA&A..42...79B,2013RPPh...76k2901B,2014ApJ...796..106J,2015MNRAS.447...49S,2016MNRAS.459.3738I,2017ApJ...836..244W,2019ApJ...880...67J,2020SSRv..216...48H,2023MNRAS.518.3036G,2024A&A...690A.106M}. 
Nevertheless, challenges remain in understanding how BHs can gravitationally capture sufficient gas from the broader environment, as investigated by \citet[e.g.][]{2023MNRAS.518.3606S}.

\textit{The intermediate-mass black hole (IMBH) seed pathway} ($10^{3-4} M_\odot$): SMBHs can grow rapidly if the seed mass is sufficiently large, such as in IMBHs. IMBHs can form through several channels, chiefly: (1) collapse of very massive stars (VMS; $10^{3-4} M_\odot$), where VMS form via runaway collisions of stars \citep{1970ApJ...162..791S,1987ApJ...319..801L,1990ApJ...356..483Q,1999A&A...348..117P,2002ApJ...576..899P,2004Natur.428..724P,2006ApJ...641..319P,2006MNRAS.368..141F,2015MNRAS.454.3150G,2016MNRAS.459.3432M,Sakurai2017,2021ApJ...908L..29G,2022MNRAS.509.3724R,2022MNRAS.515.5106W,2024MNRAS.531.3770R,2024Sci...384.1488F,2024AJ....167..191P,2025arXiv251100200P,Askar2025,GonzalezPrieto2025,Pacucci2025,Vergara2025,Vergara2026} and (2) hierarchical mergers of stellar-mass BHs via gravitational wave (GW) radiation \citep{2002MNRAS.330..232C,2004ApJ...616..221G,2009ApJ...692..917M,2015MNRAS.454.3150G,2016ApJ...831..187A,Antonini_2019,2020MNRAS.498.5652K,2021MNRAS.501.5257R,2021MNRAS.505..339M,2022MNRAS.511.5797M,2026arXiv260204176N}. Both channels are expected to occur in dense stellar systems, such as Pop III star clusters, globular clusters and galactic nuclear star clusters (NSCs). However, each channel faces challenges. For the VMS channel, it remains uncertain how VMS form via stellar mergers and how they subsequently evolve. Strong stellar winds and intense feedback, together with the possibility of general-relativistic instability supernovae, may prevent them from retaining enough mass to collapse into an IMBH \citep{2002ApJ...572L..39S,Glebbeek2009,Sakurai2015,2017PhRvD..96h3016U,2022ApJ...933..170F,2025A&A...695A.122U}. For the hierarchical merger channel, GW recoil can impart large kick velocities to the merger remnant, potentially ejecting the BH from its host system like GCs \citep{Morawski_2018}, and thereby quenching further mergers and growth. Only dense NSCs with sufficiently high escape velocities are likely to grow IMBHs and SMBHs \citep{Antonini_2019}. Even then, the IMBH formation timescale ($\ge10^3$~Myr) is too long to fully account for high-redshift SMBHs. As an alternative channel, some studies have also explored the growth of IMBHs through BH-star interactions, including tidal disruption events of unbound stars \citep{2017MNRAS.467.4180S} and direct BH-star collisions \citep{2022ApJ...929L..22R}. However, this pathway is generally considered subdominant.
    
\textit{The heavy seed pathway} ($10^{5-6} M_\odot$): in this scenario, SMBHs form from the direct collapse of gas in atomic-cooling halos \citep{2003ApJ...596...34B,2006MNRAS.370..289B,2008MNRAS.387.1649B,2010MNRAS.402..673B,2013MNRAS.436.2989L,2015ApJ...810...51M}. The gas must maintain a high temperature ($\sim5000$~K) to prevent efficient cooling by metals or molecular hydrogen (H$2$), which would otherwise lead to gas fragmentation. Under these conditions, the gas can directly collapse to a supermassive star (SMS) of $10^{5-6} M_{\odot}$, which subsequently collapses into a massive BH seed. Proposed mechanisms to achieve these conditions include, for instance, an unusually high baryon-to-dark matter streaming velocity that delays H$_2$ cooling \citep{2014MNRAS.439.1092T}. 
However, such models require extreme conditions: pristine, metal-free gas prior to the formation of first stars; a mechanism to suppress the formation of H$_2$; and a quiescent, non-fragmenting collapse \citep{2008ApJ...686..801O,2014ApJ...792...78L,2016ApJ...823...40L}. Consequently, this pathway is both physically challenging to realize and difficult to verify observationally.

Among these pathways, hierarchical BH mergers offer a comparatively simple scenario, avoiding the significant uncertainties associated with gas accretion and stellar evolution models.
However, creating an IMBH for a high-redshift SMBH seed requires rapid formation via ``runaway collisions'' of BHs, similar to the formation of VMS.
The main bottleneck is the long IMBH formation timescale: high densities are required to achieve escape velocities that retain BBH merger remnants against large GW recoils \citep{Antonini_2019}, but they also significantly delay the buildup of sufficient hard BBHs to drive GW mergers due to the low formation rate predicted by three-body scattering theory \citep{1993ApJ...403..271G,2024MNRAS.531..739G}. 

However, real $N$-body systems are more complex. Using a large suite of N-body simulations of extremely dense BH systems, we demonstrate that hard BBHs can form efficiently in dense stellar systems ($\ge10^8 M_\odot \text{pc}^{-3}$), triggering "runaway GW mergers" that build IMBHs within $10$ Myr from the formation of the BH subsystem. In such environments, hard BBHs form rapidly on $\sim10^{-2}$~Myr timescales via long-range $N$-body perturbations that generate  chains of transient binaries,
which progressively harden through exchanges until a final BBH becomes tight enough for GW-driven merger. We also identify the precise criteria for this runaway formation channel. 

Such dense stellar systems may be favored in high-redshift, low-metallicity star-forming environments, particularly feedback-free starbursts \citep[FFBs; ][]{2023MNRAS.523.3201D,2025arXiv251107578D}, where thousands of star clusters merge after wet compaction to form compact central clusters that may ultimately produce LRDs.

This paper is organized as follows: Section 2 introduces the computational methods and the star cluster models employed in this study. Section 3 presents maximum BH masses and growth from simulations, analyze the runaway BBH merger process, compares BBH merger properties with GW events such as GW190521 and GW231123 and compare with observations and star formation models. Section 4 discusses the model's limitations and future directions. Finally, Section 5 summarizes our conclusions.

\section{Method} 

\subsection{Simulation Codes}

We use \textsc{petar} \citep{Wang_2020} to perform $N$-body simulations of the dense BH system models.
\textsc{petar} is a high-performance code designed to simulate stellar systems with collisional effects, including close encounters, binaries, and hierarchical multiples. It is developed based on the Framework for Developing Particle Simulators (FDPS) \citep{2016PASJ...68...54I,71-Namekata.2018,2020PASJ...72...13I}, combining particle-tree and partilce-particle methods \citep{2011PASJ...63..881O} to resolve both long-range and short-range gravitational interactions. Long-range forces are computed using an MPI and OpenMP parallel Barnes–Hut tree algorithm \citep{1986Natur.324..446B} with a second-order leapfrog integrator, resulting in a computational cost of $O(N \log{N})$. To accurately modelling binaries, hyperbolic encounters, and hierarchical multiples, \textsc{petar} employs a fourth-order Hermite integrator with the slowdown algorithmic regularization (SDAR) method \citep{2020MNRAS.493.3398W}. 

\subsection{GW Emission And Recoil Models}

With the binary stellar evolution package \textsc{bse} \citep{72-Hurley.2000,73-Hurley.2002}, \textsc{petar} can model GW-driven mergers of compact binaries, such as BBHs, using orbit-averaged evolution based on \citet{1963PhRv..131..435P}. The average energy and angular momentum loss rates due to GW emission are given by \citep{1964PhRv..136.1224P}:
\begin{align}
\left\langle \frac{dE}{dt} \right\rangle &= 
-\frac{32}{5} \frac{G^4 m_1^2 m_2^2 (m_1 + m_2)}{c^5 a^5 (1 - e^2)^{7/2}} 
\left( 1 + \frac{73}{24} e^2 + \frac{37}{96} e^4 \right) \\
\left\langle \frac{dL}{dt} \right\rangle &= 
-\frac{32}{5} \frac{G^{7/2} m_1^2 m_2^2 (m_1 + m_2)^{1/2}}{c^5 a^{7/2} (1 - e^2)^2} 
\left( 1 + \frac{7}{8} e^2 \right)
\end{align}
where $c$ is the speed of light, $G$ is the gravitational constant, $m_1$ and $m_2$ are the masses of two of BBH, $a$ is the semi-major axis, and $e$ is the orbital eccentricity. Compared with the more accurate Post-Newtonian treatment, this orbital averaged approach combined with the SDAR method is several orders of magnitude faster in $N$-body simulations, making it well-suited for studying GW mergers in dense stellar systems. 

However, the original \textsc{petar} does not include GW recoil kicks, which are essential for studying IMBH formation via hierarchical BBH mergers. We therefore implement GW recoil kicks in \textsc{petar}, incorporating the effects of three-dimensional BH spins. We adopt the recoil prescriptions from the \textsc{precession} code \citep{2016PhRvD..93l4066G,2023PhRvD.108b4042G}, as detailed in Appendix~\ref{sec:GWkick}.

\subsection{Construction of Dense BH Cluster Models}

\begin{table}[h]
    \centering
    \scriptsize 
    \caption{Initial conditions for BH cluster simulations.}
    \label{tab:NSCs}
    \begin{tabular}{@{} c c c c c c c @{}}
        \toprule
        Name prefix & $\mbh$ distribution & $\Mbh (\pm 2\% )(M_\odot)$ &  $\rbh \, (pc)$ & $\Ms(\pm 2\% )(M_\odot)$ & $\rs(\pm 2\% )(pc)$ & simulation time $(Myr)$\\
        \midrule
        M50k-IMF-BH & IMF & 50000 & 0.1   & none & none & 352\\
                   &     &       & 0.05  & none & none & 493\\
                   &     &       & 0.01  & none & none & 208\\
                   &     &       & 0.005 & none & none & 59\\
                   &     &       & 0.001 & none & none & 13\\
        \midrule
        M50k-IMF-ALL & IMF & 50000 & 0.1   & 796306 & 0.75  & 470\\
                    &     &       & 0.05  & 741273 & 0.37  & 500\\
                    &     &       & 0.01  & 730858 & 0.073 & 103\\
                    &     &       & 0.005 & 727567 & 0.036 & 28\\
        \midrule
        M50k-SM-BH & SM & 50000 & 0.1   & none & none & 470\\
                  &    &       & 0.05  & none & none & 401\\
                  &    &       & 0.01  & none & none & 48\\
                  &    &       & 0.005 & none & none & 26\\
                  &    &       & 0.001 & none & none & 8\\
        \midrule
        M50k-SM-ALL & SM & 50000 & 0.1  & 796306 & 0.75  & 500\\
                   &    &       & 0.05 & 741273 & 0.37  & 259\\
                   &    &       & 0.01 & 730858 & 0.073 & 55\\
        \midrule
        M5k-IMF-BH & IMF & 5000 & 0.1   & none & none & 500\\
                  &     &      & 0.05  & none & none & 500\\
                  &     &      & 0.01  & none & none & 500\\
                  &     &      & 0.005 & none & none & 500\\
                  &     &      & 0.001 & none & none & 319\\
        \midrule
        M5k-IMF-ALL & IMF & 5000 & 0.1   & 79630 & 1     & 496\\
                   &     &      & 0.05  & 74127 & 0.5   & 500\\
                   &     &      & 0.01  & 73085 & 0.1   & 494\\
                   &     &      & 0.005 & 72756 & 0.049 & 500\\
                   &     &      & 0.001 & 69518 & 0.01  & 500\\
        \midrule
        M5k-SM-BH & SM & 5000 & 0.1   & none & none & 360\\
                 &    &      & 0.05  & none & none & 496\\
                 &    &      & 0.01  & none & none & 500\\
                 &    &      & 0.005 & none & none & 500\\
                 &    &      & 0.001 & none & none & 264\\
        \midrule
        M5k-SM-ALL & SM & 5000 & 0.1   & 79630 & 0.85  & 500\\
                  &    &      & 0.05  & 74127 & 0.42  & 327\\
                  &    &      & 0.01  & 73085 & 0.08  & 459\\
                  &    &      & 0.005 & 72756 & 0.042 & 474\\
                  &    &      & 0.001 & 69518 & 0.008 & 191\\
        \bottomrule
    \end{tabular}
    \begin{itemize}
        \item Note: the full model name consists of a prefix and $\rbh$. For example, M50k-IMF-BH-r0.1 denotes the model M50k-IMF-BH with $\rbh=0.1$~pc. Because the model M50k-SM-BH-r0.001 runs more slowly, it was simulated only 8 times, whereas all other models were simulated 45 times.
    \end{itemize}
\end{table}

We assume the BH system is embedded within a massive ($\sim10^7$ stars) dense cluster, such as a NSC or a massive GC. Since star-by-star $N$-body simulations of such systems are very time-consuming. As our study focuses on BBH mergers rather than stellar interactions, we model the surrounding halo stars as a smooth Plummer potential \citep{1911MNRAS..71..460P}. This approach significantly reduces computational cost, allowing us to simulate extremely dense BH systems across multiple models with varied initial conditions and random seeds, although the full suite still required about 600,000 CPU hours on the Ningxia NC-E supercomputer. These simulations provide the statistics necessary to investigate how IMBH formation rate depends on cluster properities.


Numerous studies have investigated BH mergers in NSCs through numerical simulations. The semi-analytic models of \citet{2016ApJ...831..187A} showed that NSCs with total masses of $\sim10^{6}\,M_\odot$, total BH masses of $\sim10^{4}\,M_\odot$, and half-mass radii of $\gtrsim 3 \,\mathrm{pc}$ can support substantial BH growth through repeated mergers. Similarly, \citet{2022ApJ...933..170F} considered NSCs of $10^{6}$--$10^{8}\,M_{\odot}$ with typical half-mass radii of $\sim 1\,\mathrm{pc}$ and found that hierarchical mergers can build a $10^3\,M_\odot$ IMBH within 1 Gyr. However, IMBH formation in these models remains too slow, and our direct $N$-body test simulations under comparable density conditions confirm that rapid IMBH formation is not achievable. We therefore explore denser initial conditions by varying the half-mass radius to identify the regime in which rapid IMBH formation can occur.

\subsubsection{BH Subsystem}


For the BH subsystem, we adopted a simple Plummer model and conduct simulations with different initial conditions, as summarized in Table \ref{tab:NSCs}, to investigate which conditions allow IMBH formation. The M50k-SM-BH-r0.001 models are computationally expensive and were therefore simulated only 8 times, while all other models were simulated 45 times. We vary the BH subsystem half-mass radius $\rbh$ (0.001, 0.005, 0.01, 0.05, 0.1 pc), and total BH mass $M_\bullet$ (5000 and 50000 $M_\odot$), using two mass spectra. The ``SM" model assumes a uniform mass of 10 $M_{\odot}$. The ``IMF" model derives BH masses from a \cite{2001MNRAS.322..231K} stellar initial mass function ($25-150~M_{\odot}$) evolved for 100 Myr using the \textsc{sse/bse} module in \textsc{petar} \citep{72-Hurley.2000,73-Hurley.2002,74-Banerjee.2020} with a metallicity $\log(Z/Z_\odot)=-1.3$. We retain only BHs and incorporate supernova kick velocities into their initial velocities. The adopted metallicity is characteristic of typical GC environments. For lower metallicities, massive stars can produce systematically more massive BHs, making rapid IMBH formation even more favorable than under our current conditions. All BH spins are initialized to zero.

\subsubsection{Halo Star Potiential}

We perform simulations both with and without stellar halo, distinguished by model names containing ``All'' and ``BH'', respectively. The stellar halo is approximated as a gravitational potential. In reality, two-body interactions between stars and BHs would provide dynamical friction that more effectively prevents BHs from escaping, while stellar wind mass loss would decrease the stellar potential and cause the system to expand. Because implementing a self-consistent time-dependent stellar potential is challenging, we instead adopt a limiting case that neglects stellar interactions and assumes a static gravitational potential.

Since we assume the BH subsystem has already formed and mass segregated within the dense star clusters, we need to generate the BH subsystem and stellar halo as two Plummer components with different half-mass radii in virial equilbrium. Simply creating and combining two independent Plummer models would violate equilibrium and lead to collapse or dispersal. To construct a balanced two-Plummer system, we use the \textsc{twoplummer} code (private communication with Holger Baumgardt) to generate equilibrium initial conditions. 

The total mass of stellar halo $\Ms$ is estimated based on the BH mass fraction, assuming the Kroupa IMF with stellar masses between $0.08-150~M_\odot$ evolved for 100~Myr with $\log(Z/Z_\odot)=-1.3$. To determine the half-mass radii of stellar halos  $\rs$, we adopt the following approach. According to \citet{2013MNRAS.432.2779B}, in star clusters hosting a significant BH population, energy generation is dominated by few-body encounters within a centrally concentrated BH subsystem. This relationship implies that the energy exchange between the stellar halo and the BH subsystem can be used to constrain $\rs$. Following this methodology, we compute $\rs$ using Equation (18) from \citep{2020MNRAS.491.2413W}:
\begin{align}
    \left(\frac{\rbh}{\rs}\right)^{5/2} = k_1 \left(\frac{\Mbh}{M}\right)^{1/2}\left(\frac{\mbhave}{\mave}\right) ^{2-\gamma} \frac{1}{1+\frac{\Ms}{\Mbh}\left(\frac{\msave}{\mbhave}\right)^{\gamma-1}} \frac{\ln \Lambda_\bullet}{\ln \Lambda}
\end{align} 
where $\rbh$ and $\rs$ denote the half-mass radii of the BH subsystem and the stellar halo, respectively. $\Mbh$, $\Ms$, and $M$ are the total masses of the BH subsystem, the stellar halo, and the entire system. $\mbhave$, $\msave$, and $\mave$ represent their corresponding mean particle masses. The Coulomb logarithm is taken as $\ln\Lambda = 0.02N$, following \citet{1996MNRAS.279.1037G}. The parameters $k_1$ and $\gamma$ are adopted from Figures~4 and~6 of \citet{2020MNRAS.491.2413W}. Then, the stellar halo potential is included in simulations using the package \textsc{galpy} \citep{2015ApJS..216...29B}. 

\section{Results}

\subsection{IMBH Formation}

\begin{figure}[h]
  \centering
  \includegraphics[width=1\textwidth]{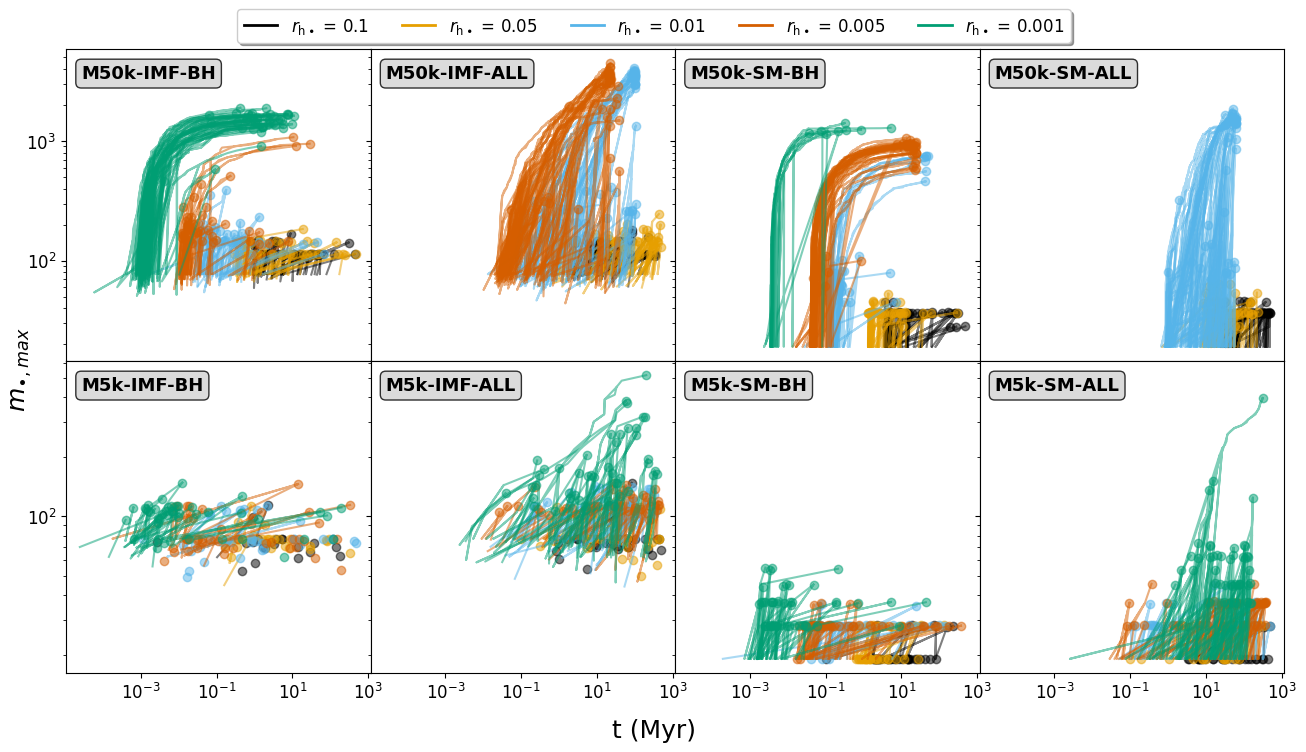}
  \caption{The maximum BH mass ($\mbhmax$) as a function of time ($t$). Model initial conditions are listed in Table~\ref{tab:NSCs}, and different colors represent different $\rbh$ values in pc.}
  \label{fig:generate_mass}
\end{figure}


To investigate how IMBH mass evolution depends on the initial conditions, we present the evolutionary curves of the system's maximum BH mass ($\mbhmax$) over time ($t$) in Figure \ref{fig:generate_mass}. IMBHs can form within 10 Myr from the formation of the BH subsystem. Specifically, an initial mass of $5 \times 10^4 \, M_\odot$ is necessary. Models such as M50k-SM-BH-r0.001, M50k-IMF-BH-r0.001, and M50k-IMF-ALL-r0.005/r0.01 all exhibit a high probability of IMBH formation. The half-mass  initial density of these models ranges from $ 5\times10^9~M_\odot/\text{pc}^3$ to $ 10^{13}~M_\odot/\text{pc}^3$. 
In the M50k-IMF-All-r0.01 model, for instance, a half-mass average density of $5 \times10^9~M_{\odot}/\text{pc}^3$ yields an IMBH of approximately 4,000~$M_{\odot}$. 
Higher densities or smaller $\rbh$ lead to shorter formation timescales.

Our simulations begin after mass segregation and the formation of the BH subsystem. The BH mass-segregation time, $T_{\rm ms,\bullet}$ \citep{1987degc.book.....S}, is 
\begin{align}
T_{\rm rh,\star} &= 0.138 \frac{N_\star^{1/2} R_{\rm h,\star}^{3/2}}{\overline{m_\star}^{1/2} G^{1/2} \ln \Lambda_{\star}}, \\
T_{\rm ms,\bullet} &= \frac{\overline{m_\star} }{\overline{m_\bullet}} T_{\rm rh,\star},
\end{align}
where $T_{\rm rh,\star}$ \citep{1987degc.book.....S} is the relaxation time of stars. For the number of stars $N_\star \approx 1.6\times10^{6}$, the stellar half-mass radius $R_{\rm h,\star} \approx 0.073$ pc, the average stellar mass $\overline{m_\star} \approx 0.465~M_\odot$ and $\mbhave \approx 24.91 M_\odot$, we obtain $T_{\rm ms,\bullet}\approx 0.14$~Myr. Including the 4-10 Myr required for main-sequence stars to form BHs, IMBHs can form in these systems within 20 Myr after star formation.

$\Mbh$ is crucial for IMBH formation. When $\Mbh$ (or the total number of BHs) is too low, effective BBH mergers become unlikely. For instance, although  the M5k-IMF-All-r0.001 model has a higher initial density due to its smaller $\rbh$, the M50k-IMF-All-r0.01 model reaches a larger  $\mbhmax$ earlier and faster. Furthermore, in M50k-IMF-All models with $\rbh=$ 0.01 and 0.005 pc, $\mbhmax$ appears to saturate after a certain value. This suggests a correlation between $\Mbh$ and $\mbhmax$. A similar saturation occurs in the VMS channel for IMBH formation, as shown in  panel~(a) of Fig.~1 in \citet{Askar2025}. This behavior arises because IMBH formation reduces the core density, making it more difficult to capture BHs and form hard binaries that can merge. 

Secondly, comparing the M50k-IMF-BH and M50k-SM-BH models with $\rbh = 0.005-0.01$~pc, where IMBHs form stochastically, shows that the SM model with equal-mass BHs has a higher probability of forming an IMBH. As indicated by Equation \ref{eq:vm}, equal-mass BHs produce smaller recoil velocities due to mass symmetry, making post-merger ejection from the system less likely. 

Finally, comparing the M50k-IMF-BH and M50k-IMF-All models reveals that existence of stellar halo potential leads to more massive IMBHs. 
However, the onset of rapid growth is more stochastic and systematically delayed for the IMF models, especially in M50k-IMF-ALL-r0.005. This delay arises from the larger velocity dispersion ($\sigma$), which increases the formation timescale of tight BBHs that eventually merge, as discussed in the next section. 

Table \ref{tab:IMBHs} summarizes the fraction of models forming IMBHs ($f_{\rm IMBH}$), the maximum IMBH mass $m_{\bullet,\rm IM,max}$, its mean $\overline{m_{\bullet,\rm IM,max}}$, the mean formation time ($\overline{t}$) and the formation time range for models that produce IMBHs.
Under identical conditions, SM models yield smaller IMBHs than IMF models. The presence of a stellar halo enhances both $f_{\rm IMBH}$ and $m_{\bullet,\rm IM,max}$ and decreases $\overline{t}$. In all models, the formation time decreases as the initial radius decreases.
\begin{table}[h]
    \centering
    \caption{IMBH formation for models}
    \label{tab:IMBHs}
        \begin{tabular}{@{} l c c c c c @{}}  
        \toprule
        Name prefix & $f_{\rm IMBH}$ & $m_{\bullet,\rm IM,max} (M_\odot)$ &  $\overline{m_{\bullet,\rm IM,max}} (M_\odot)$& $\overline{t}$ (Myr) & Time range (Myr)\\
        \midrule
        M50k-IMF-BH-r0.005 &  1/45  & 1075.5182 & 1075.5182& 3.8024 & [3.8024,3.8024]\\
        M50k-IMF-BH-r0.001 &  40/45  & 1890.5832 & 1505.6104& 0.0158  & [0.0032,0.0691]\\
        M50k-IMF-ALL-r0.01 &  22/45  & 4063.9096 & 3366.0509& 10.8186 & [1.1972,94.8868] \\
        M50k-IMF-ALL-r0.005 &  36/45  &4470.2936&  2965.2926 & 2.9354 & [0.1937,20.3276] \\
        M50k-SM-BH-r0.005 & 1/45 & 1061.6314&1061.6314& 3.2637 & [3.2637,3.2637]\\
        M50k-SM-BH-r0.001 & 8/8 & 1410.3802& 1236.0859& 0.0241  &  [0.0145,0.0380]\\
        M50k-SM-ALL-r0.01 & 5/45 &1282.4926&1127.1320&11.9525  & [8.3317,15.0040] \\
        \bottomrule
        \end{tabular}\\

    \begin{itemize}
        \item Note: $f_{\rm IMBH}$ is defined as the ratio of the number of simulations in which the model produces an IMBH to the total number of simulations. If no IMBH is formed in any of the simulations, the model is not included in the Table.

    \end{itemize}
\end{table}

\subsection{Early BBH Formation and Mergers}

\begin{figure}[h]
  \centering
  \includegraphics[width=1\textwidth]{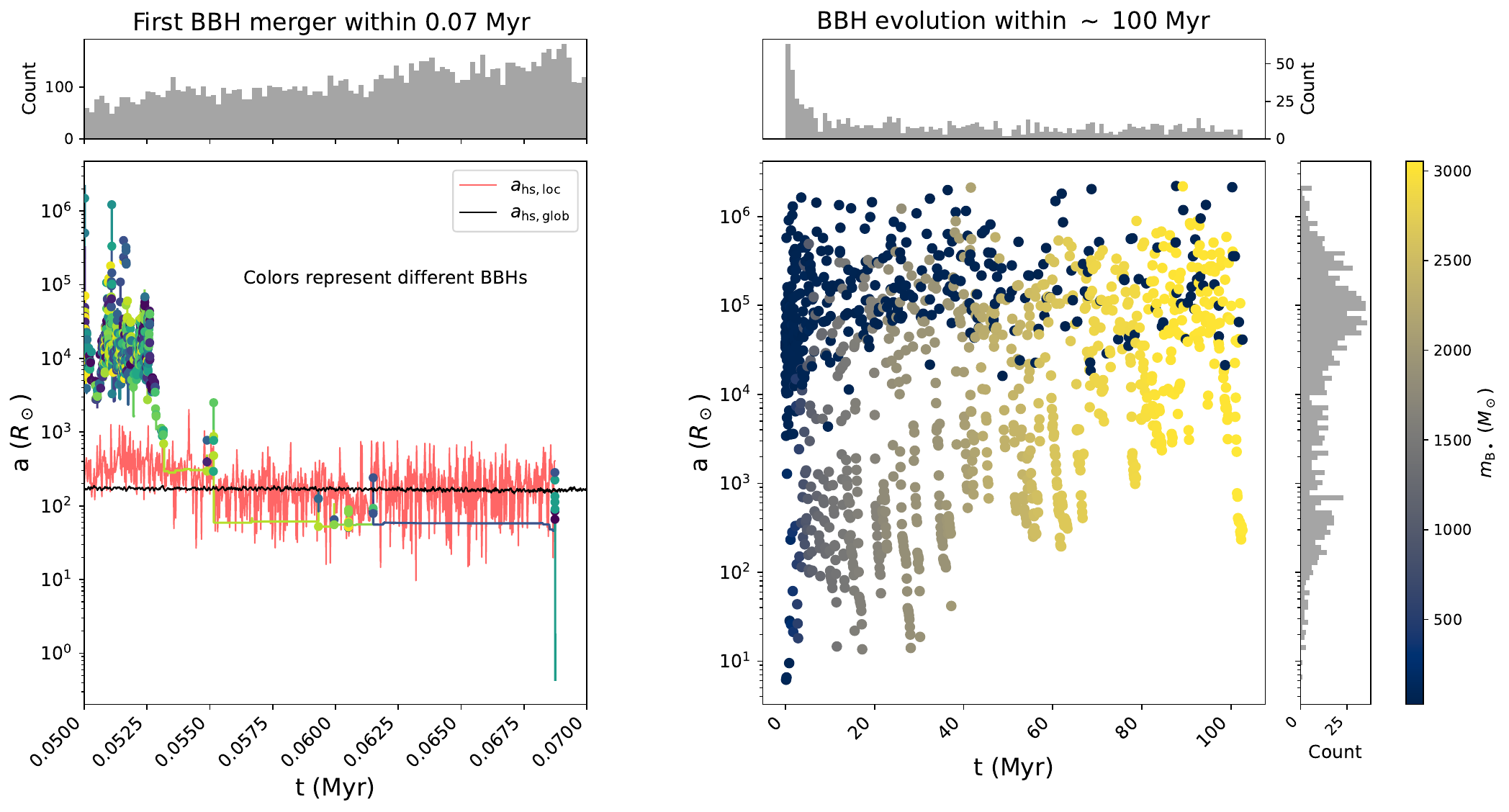}
  \caption{ Semi-major axis ($a$) of BBHs as a function of time $t$ for the M50k-IMF-All-r0.01 model. The left panel adopts a sampling interval of $2\times10^{-7}$ Myr, revealing the detailed early evolution of the system between 0.05 and 0.07 Myr. The lines with points denote the various transient BBHs that appear throughout the evolutionary history of a particular hard BBH that merges at 0.06874 Myr. $a_{\rm hs,glob}$ and $a_{\rm hs,loc}$ denote the hard-soft binary boundaries for the global and local regions, respectively. The histograms at the top represent the number of distinct BBHs in each time bin ($2\times10^{-7}$ Myr). The right panel employs a sampling interval of 0.01 Myr, revealing BBH evolution within $\sim 100$ Myr, with the color of each point indicating the total mass of the BBH system ($M_{\rm BBH}$). The histograms at the top and right represent the number of distinct BBHs in each time bin (0.01 Myr) and the marginal distribution of BBHs along the semi-major axis $a$, respectively. 
  }
  \label{fig:perigee}
\end{figure}

\begin{figure}[h]
    \centering
    \includegraphics[width=0.8\textwidth]{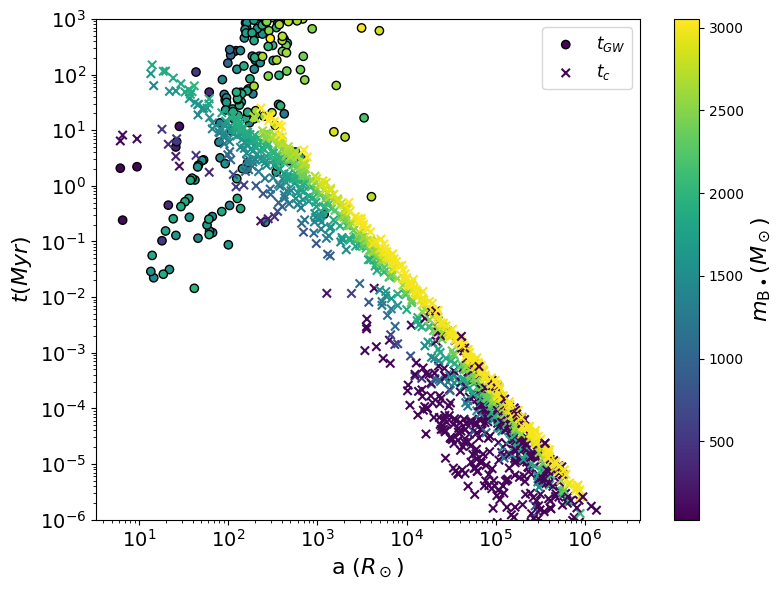}
    \caption{The relationship between the semi-major axis $a$ and evolution timescales for BBH mergers in the M50k-IMF-All-r0.01 model. $t_{\rm GW}$ represents by dots, denotes the GW merger timescale, while $t_{\rm c}$ shown as crosses, denotes the collisional (close-encounter) timescale. }
    \label{fig:cluster2}
\end{figure}

To investigate why IMBHs can form within 10~Myr from the formation of the BH subsystem in dense systems, we analyze BBH formation in the M50k-IMF-All-r0.01 model. Figure~\ref{fig:perigee} presents the evolution of the BBHs' semi-major axis ($a$) over time $t$. The left panel uses a sampling interval of $2 \times 10^{-7}$ Myr to resolve the rapid early evolution of transient BBHs about the earliest BBH merger, while the right panel uses a sampling interval of 0.01~Myr to show the long-term evolution of all BBHs in the system within $\sim100$ Myr. Here we clarify two definitions of the hard-soft binary boundary $a_{\rm hs} = (G m_1 m_2)/(2 \overline{m_\bullet} \sigma^2)$ \citep{Heggie1975,Hills1975} used in the subsequent analysis, where $\sigma$ is velocity dispersion: the global boundary, computed from averaged quantities within $\rbh$, and the local boundary, calculated based on the immediate environment of the binary.

In the early phase ($ 0.5\,\mathrm{Myr} \leq t \lesssim 0.7\,\mathrm{Myr}$; left panel), many globally soft BBHs form rapidly but are quickly disrupted.
Rarely, a globally hard BBH forms and merges via GW emission at $t=0.06874$~Myr. 
We trace this BBH's evolution (colored points marking different BBHs along the $a$ track) and find that it does not arise from a single three-body event; instead, it emerges through repeated binary exchanges. It originates from globally soft BBHs with $a\sim10^4-10^6~R_\odot$ at $0.05-0.0525$~Myr. Subsequent close encounters drive frequent component swaps, and along one exchange branch $a$ is inherited and progressively hardened. The hardening is stepwise rather than continuous, with a few strong encounters causing abrupt drops in $a$. A strong encounter then produces a hard binary that crosses the globally hard-soft boundary ($\sim 400~R_\odot$ in this model) at 0.0525-0.0550~Myr. After a brief stable phase, further strong interactions and exchanges results in a highly eccentric hard BBH that enters the GW dominated regime and rapidly coalesces at 0.06874~Myr. 

This result provides a new perspective on hard-binary formation, suggesting that early BBH mergers are driven by cumulative dynamical exchanges rather than isolated hardening events.
This pathway allows globally soft binaries to transition into globally hard ones, seemingly contradicting the classical hard-soft binary theory, which predicts that soft binaries become softer. However, in star clusters the hard-soft boundary fluctuates locally: a binary that is soft in one region may be hard in another, enabling hardening through close encounters. In the left panel of Figure \ref{fig:perigee}, the red curve shows the hard-soft boundary ($a_{\mathrm{hs,loc}}$) calculated from the 5 nearest surrounding BHs for transient BBHs in the hard-BBH formation chain. At $t = 0.0530$ Myr, the local boundary lies above the global hard-soft boundary ($a_{\mathrm{hs,glob}}$, red line), allowing BBHs that are locally hard but globally soft to further harden during strong dynamical encounters and eventually cross the global hard-soft boundary.
Although such binaries may later be disrupted in softer regions, the hardening gained in harder regions can be inherited through binary exchanges. Repeated exchanges accumulate this hardening, allowing binaries to cross the global hard-soft boundary until one becomes sufficiently stable and eventually merges.

Variations in the local hard-soft boundary arise primarily from changes in the velocity dispersion, which fluctuates between $30$ and $150\,\mathrm{km\,s}^{-1}$. Although velocity dispersion also increases on longer timescales due to cluster evolution, this occurs after IMBH formation ($2-3$ Myr) and therefore plays only a minor role during the first 0.01~Myr evolution considered here. 

This rapid formation of hard BBHs and repeated mergers ultimately grows an IMBH within 10~Myr. As shown in the right panel and its top histogram of Figure~\ref{fig:perigee}, the merger rate is much higher during first 10~Myr growth phase than at later times, when mergers are dominated by the IMBH with low-mass BHs. 


This hard-BBH formation channel is much faster than predicted by classical three-body scattering theory. 
The cumulative three-body formation rate of hard BBHs is \citep{1993ApJ...403..271G,2024MNRAS.531..739G,atallah2024binaryformationinitiallyunbound} 
\begin{align}
\Gamma_\mathrm{F}(a<a_{\mathrm{hs}}, \chi_{\mathrm{IV}}) 
\approx 0.081 \,
\frac{G^5 m^5 n^3}{\sigma^9}
(\frac{a_{\mathrm{hs}}}{ a_{\mathrm{fs}}})^3
\chi_{\mathrm{IV}}^{1/2}
\label{eq:hardbinary}
\end{align}
where $n$ is the number density, $m$ is the mass of a single BH, $a_{\mathrm{fs}}$ is is the boundary defined by the fast and slow criteria proposed by \citep{1990AJ.....99..979H}, and $\chi_{\mathrm{IV}}$ controls the characteristic strength. This expression implies a steep suppression of BBH formation with increasing $\sigma$.

For the BH-subsystem core over $0.05-0.07$~Myr, using mean values 
$m \approx 35.02\,\mathrm{M_\odot}$, 
$n \approx 2.28 \times 10^{10}\,\mathrm{pc^{-3}}$, 
$\sigma \approx 123.89\,\mathrm{pc\,Myr^{-1}}$, 
$a_{\mathrm{hs}} \approx 5.22 \times 10^{-6}\,\mathrm{pc}$,
$a_{\mathrm{fs}} \approx 1.04 \times 10^{-5}\,\mathrm{pc}$
and $\chi_{\mathrm{IV}} = 10$ (It is a close match to the prediction of the hard binary formation from \citet{1993ApJ...403..271G}), we obtain $\Gamma_{\mathrm{F}} \approx 0.033 \,\mathrm{Myr^{-1}}$ within the core radius $r_\mathrm{c} \approx 8.57 \times 10^{-4}\,\mathrm{pc}$. This is far below our numerical results, which produce the first hard-BBH merger by $\sim0.07$ Myr and followed by many more at later times, with an average of 16 mergers occurring within the first 1 Myr. 


In contrast, the soft-binary formation rate in our simulations is consistent with the three-body scatter theory \citep{atallah2024binaryformationinitiallyunbound}:
\begin{align}
\Gamma_{\mathrm{F}}(a > r_{\mathrm{IV}}) 
\approx 16.8 \,
\frac{G^2 m^2 n^3}{\sigma^3}
r_{\mathrm{IV}}^3 
\label{eq:softbinary}
\end{align}
where $r_{\mathrm{IV}} = 2~a_{\mathrm{fs}}~\chi_{\mathrm{IV}}$ denotes the radius of the three-body interaction volume.
In our simulations, taking $r_{\mathrm{IV}} = 20~a_{\mathrm{fs}}$ 
and adopting the same physical conditions as for hard binaries yields 
$\sim 1.4 \times 10^5\,\mathrm{Myr^{-1}}$, similar to our simulation result. However, this formula is derived from curve fitting of simulation data. Its cumulative probability range is constrained by the value of $\chi_\mathrm{IV}$, failing to cover the entire parameter space of $a$. Consequently, we only treat it as a baseline reference in our study. 

Furthermore, Following \citet{2020PhRvD.101l3010S}, the GW capture rate between single BHs $\Gamma_{\mathrm{ss}}$ is:
\begin{align}
\mathcal{R}_{\mathrm{ss}} &= \left( \frac{85\pi}{24\sqrt{2}} \right)^{2/7} \mathcal{R}_m \left( \frac{c^2}{v^2} \right)^{2/7}, \label{eq:Rss} \\
\Gamma_{\mathrm{ss}} &= \frac{8\pi^2 G}{\mbhave} \int_0^\infty \frac{\mathcal{R}_{\mathrm{ss}} \rho^2}{v} r^2 dr. \label{eq:ratess}
\end{align}
Here $\mathcal{R}_{\rm m}$ is the Schwarzschild radius of a BH, $R_{\mathrm{ss}}$ is the single-single GW capture cross section, and $v$ and $\rho$ are the mean velocity and mass density of the BH subsystem at radius $r$. Using $\Mbh =50,000 \, M_\odot$, $\rbh =0.01 \, \text{pc}$, and $\mbhave =35.02 \, M_\odot$, and estimating $\rho$ and $v$ from a Plummer model, we find a single-single GW capture rate of order $0.64 \, \mathrm{Myr}^{-1}$. This formation rate, however, still remains below the BBH merger rate observed in our simulations.



Overall, our results suggest a new channel for rapid IMBH formation: Local fluctuations in spatial density enable "globally soft" BBHs to survive and harden into stable, hard binaries. Although rare, these mergers can seed subsequent hierarical growth, producing an IMBH within 10~Myr from the formation of the BH subsystem despite the low hard-BBH formation rate predicted by Equation~(\ref{eq:hardbinary}). 

\subsection{IMBH-dominated BBH Formation and Mergers}

While early hard BBH formation is driven by stochastic interactions among stellar-mass BHs, the evolution after 10~Myr is dominated by the IMBH. The right-hand panel of Figure \ref{fig:perigee} displays BBHs recorded in snapshots from 0-100~Myr at interval of 0.1~Myr. Although soft stellar-mass BBHs can form, they are quickly disrupted and cannot evolve into hard binaries. In contrast, only BBHs involving the IMBH can harden and eventually merge, allowing the IMBH to grow continuously to about 3000~$M_\odot$.

During the IMBH-dominated stage, the histogram on the right-hand panel in Figure~\ref{fig:perigee} shows the cumulative number of BBHs for different values of $a$. The distribution of $a$ exhibits two distinct peaks. The first, at $a\sim10^5\,R_\odot$, corresponds to soft, dynamically formed binaries that are predominantly disrupted. The second peak, at $a\sim4\times10^2\,R_\odot$, marks a transition in the dominant hardening mechanism，we can also see that it is parallel to the boundary between fast and slow binaries. At large separations, binary evolution is driven primarily by dynamical encounters; as $a$ decreases, the encounter timescale increases, temporarily slowing the orbital decay. Once $a$ becomes sufficiently small, GW emission dominates and rapidly drives the binary toward merger.

Figure \ref{fig:cluster2} shows the collisional timescale $\tcoll$, which represents the close-encounter timescale between BBHs and intruders, and the GW merger timescale $\tGW$ dependence on the semi-major axis $a$. 
According to the expression of \citet[Eq. 7.194]{1987gady.book.....B}, the formula about $\tcoll$ is
\begin{align}
    \frac{1}{\tcoll} = 4 \sqrt{\pi}n\sigma\left(r^2_{\rm coll}+\frac{Gm}{\sigma^2} r_{\rm coll}\right),
    \label{eq:tcoll}
\end{align}
where $r_{\mathrm{ coll}}$ denotes the pericenter distance of the BBH.
And the GW merger timescale $\tGW$. The formula for $\tGW$ is taken from \citet{1964PhRv..136.1224P}:
\begin{align}
    c_0 &= \frac{a(1-e^2)}{e^{12/19}} \frac{1}{(1+\frac{121}{304}e^2)^{870/2299}} \\
    \beta &= \frac{64}{5} \frac{G^3m_1m_2(m_1+m_2)}{c^5}\\
    \tGW &= \frac{12}{19}\frac{c_0^4}{\beta} \times \int_{0}^{e}\frac{dee^{29/19}[1+(121/304)e^2]^{1181/2299}}{(1-e^2)^{3/2}}
    \label{eq:tgw}
\end{align}

The two timescales intersect at $a\sim10^2$--$10^3\,R_\odot$, consistent with the location of the second peak in Figure~\ref{fig:perigee}. Below this threshold, $\tGW < \tcoll$, implying that binaries merge via GW emission before subsequent close encounters can disrupt them.


\subsection{BBH Merger Property and Gravitational Wave Events}

Hierarchical BBH mergers leading to IMBH formation can produce GW events potentially detectable by next-generation space-based detectors like LISA \citep{2024arXiv240207571C}, TianQin \citep{2016CQGra..33c5010L,2020CQGra..37r5013L}and Taiji \citep{2017NSRev...4..685H} and the third-generation ground-based gravitational-wave observatory like Einstein Telescope \citep{2026JCAP...03..081A}. 
We analyze the properties of individual BBH mergers from our models, including the resulting BH spin $\chi$, kick velocity $\vk$, and pre-merger orbital parameters such as mass ratio $q$.
\subsubsection{Spin and Kick Velocity}


\begin{figure}[h]
  \centering
  \includegraphics[width=1\textwidth]{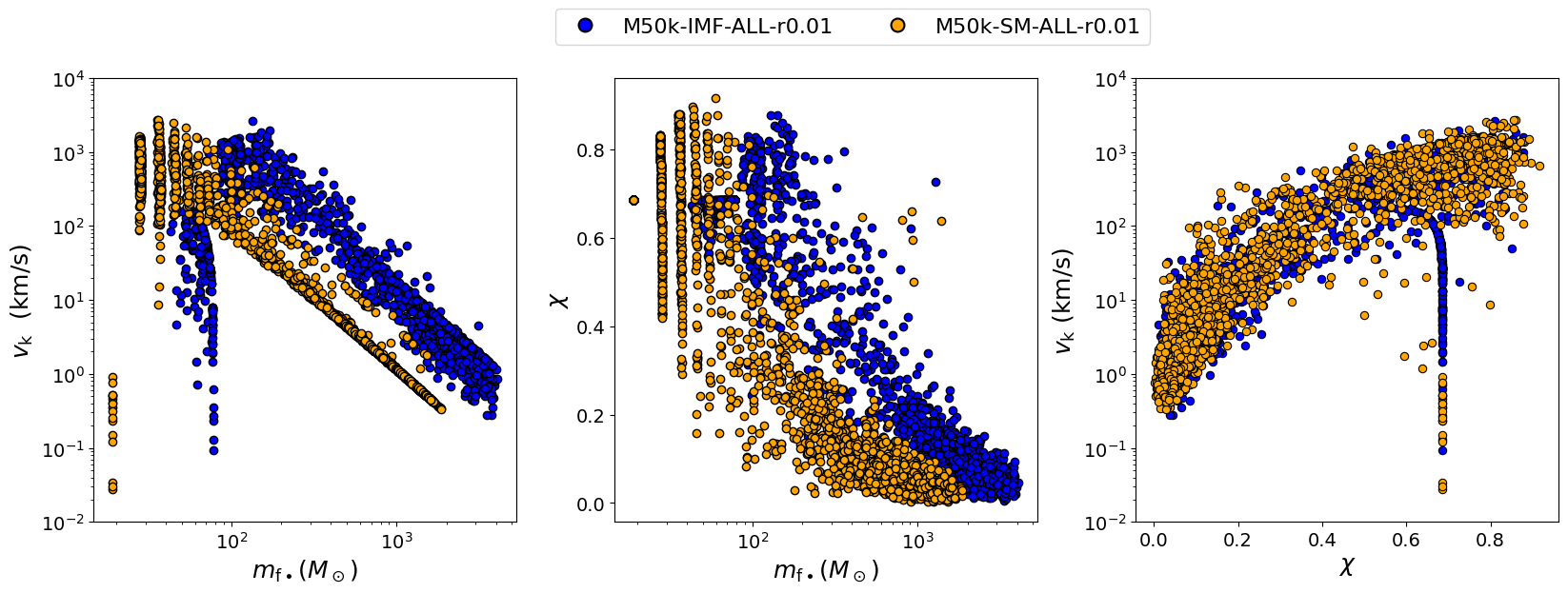}
  \caption{The relationships between post-merger BH mass $\mbhf$ and GW kick velocity $\vk$ (left), spin $\chi$ (middle) and between $\chi$ and $\vk$ (right) are shown for the M50k-IMF-All-r0.01 and M50k-SM-All-r0.01 models.  
  }
  
  \label{fig:dmsr}
\end{figure}

The BH $\vk$ is crucial in determining whether it can undergo subsequent mergers. A sufficiently low $\vk$ allows the BH to remain bound to its host system, enabling the formation of new BBHs and further mergers. Here, we present the relationships between post-merger BH mass $\mbhf$ and kick velocity $\vk$, spin $\chi$ and between $\chi$ and $\vk$
in Figure \ref{fig:dmsr}, which shows that as $\mbhf$ increases, both $\vk$ and $\chi$ first increase and then decrease (with the initial spin set to zero), and the kick velocity $\vk$ shows a positive correlation with the dimensionless spin $\chi$. Since our initial spins are zero and the initial masses are similar, the first merger for each BH can produce a low kick velocity while generating a large final spin, primarily from the orbital angular momentum. This phenomenon is more pronounced in the SM model due to its identical initial masses. This explains both the low-mass velocity tail in the velocity-mass diagram and the continous curve with a wide range of $\vk$ and $\chi=0.7$ in the velocity-spin diagram. 


Comparing the IMF and SM models, the IMF model have systematically higher $\vk$ and $\chi$ at a given $\mbhf$, resulting from unequal BBH component masses. Equal-mass BBHs lack mass asymmetry and consequently produce lower $\vk$ and $\chi$.


Because our simulations are gas-free and high-density, BH spins are expected to be isotropically distributed. During hierarchical mergers, retrograde orbits are more efficient at spinning down BHs, as their last stale orbits occur at larger radii and therefore carry more angular momentum than prograde orbits\citep{2003ApJ...585L.101H,2007arXiv0707.0711M,2013ApJ...775...94V,2026arXiv260204176N}. As a result, BH spins tend to decrease progressively through successive mergers. Since the distribution of spin orientations is random, $\vk$ is primarily influenced by the magnitude of $\chi$. Consequently, $\vk$ driven by $\chi$ generally decreases with increasing remnant mass. This produces a characteristic trend in which both $\vk$ and $\chi$ rise at early stages, then decline at higher masses.

\subsubsection{BBH Orbital Parameters}

%
\begin{figure}[h]
  \centering
  \includegraphics[width=1\textwidth]{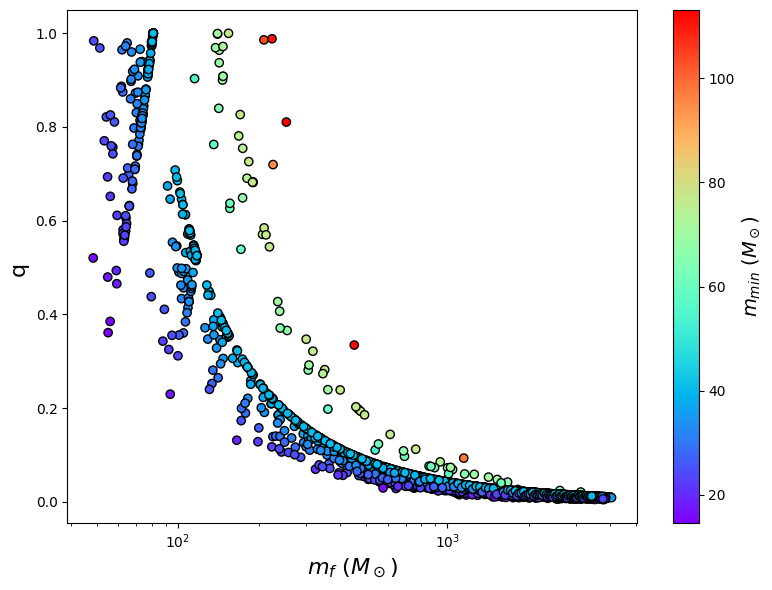}
  \caption{Relation between post-merger BH mass ($\mbhf$) and component mass ratio ($q$) of pre-merger BBHs for the M50k-IMF-ALL-r0.01 model. Colors represent the smaller BH mass $\mbhmin$ in BBHs.}
  \label{fig:BHp}
\end{figure}

The panel of Figure \ref{fig:BHp} show the mass ratio ($q$) of pre-merger BBH components versus the post-merger BH mass ($\mbhf$) for the M50k-IMF-ALL-r0.01 model. Two main trends are evident. First, $\mbhf$ and $q$ exhibit an anti-correlation, forming discrete curves determined by the smaller component mass ($\mbhmin$), with the primary curve corresponding to $\mbhmin=40.5~M_\odot$. This pattern reflects the sequential mergers of an IMBH growing through BBH mergers. Second, along the main curve with $\mbhmin= 40.5~M_\odot$, several straight lines with a positive correlation between $\mbhf$ and $q$ appear. They originate from BBH mergers involving one massive BH of $40.5~M_\odot$ (or its multiples) and another BH of varying mass. These features can be explained by the stellar evolution model and IMF setup, where many massive stars evolve into $40.5~M_\odot$ BHs below the pair-instability supernova gap, making mergers with such BHs frequent. The resulting merger remnants then produce both the discrete IMBH-BBH merger curves and the sequences of straight lines from lower-mass mergers. 

\subsection{Events rate of GW190521- and GW231123-like events}

Recently, the detection of the GW event GW190521 ($m_1 = 85^{+21}_{-14} M_\odot,m_2 = 66^{+17}_{-18} M_\odot, \chi_1=0.69^{+0.27}_{-0.62}, \chi_2=0.73^{+0.24}_{-0.64}$) \citep{2020PhRvL.125j1102A, 2020ApJ...900L..13A, 2022A&A...659A..84A} and GW231123 ($m_1 = 137^{+23}_{-18} M_\odot,m_2 = 101^{+22}_{-50} M_\odot, \chi_1=0.90^{+0.10}_{-0.19}, \chi_2=0.80^{+0.20}_{-0.52}$) \citep{2025ApJ...993L..25A} has established the existence of a BH population that cannot form directly through pair-instability supernovae within this mass range; these are known as pair-instability mass gap BHs (PIBHs). In M50k-IMF-ALL-r0.01 models within 100~Myr, PIBHs can form efficiently. In models with IMBHs the PIBH formation rate via BBH mergers is 0.7926-11.7724 per Myr per cluster. In models without IMBHs, the mean rate decreases to 0.3607-2.0386 per Myr per cluster. The data presented are from 45 independent realisations of the M50k-IMF-ALL-r0.01 model. The upper and lower bounds shown reflect the statistical spread across these 45 simulations. The fractions reported correspond to the mass of merger remnants that fall within the PIBH mass range. And these statistics represent averages over the full duration of our simulations, not just the first 10 Myr. However, the PIBH formation rate drops sharply after the initial 10 Myr. These PIBHs can subsequently form new BBHs and produce GW190521- and GW231123-like events.

Table \ref{tab:GWs} compares the mean counts per Gyr of GW190521-like and GW231123-like events for all radial bins of the M50k-IMF-BH and M50k-IMF-ALL models, in cases with ($\Gamma_{\rm I}$) and without ($\Gamma_{\rm n}$) an IMBH, along with their fractional contribution ($P$) among all BBH mergers. The observed mass and spin uncertainties of these two events are adopted as the selection criteria. We restrict the analysis to the time range of $0-10$~Myr, and only include models with runtime approaching or exceeding 10~Myr in the calculation. And we divide them into cases where the entire model produces IMBHs and cases where it does not produce IMBHs. Based on these two classifications, we obtain the internal merger rate for each simulation, and average them to obtain $\Gamma_{\rm I}$ and $\Gamma_{\rm n}$. The SM model is discounted because its uniform masses lead to merger products with nearly identical binary masses. The M5k models are excluded from the table due to the insufficient number of relevant events they produced. 
Two samples are considered: events matching the mass range only, and those matching both mass and spin ranges. 

$\Gamma_{\rm I}$ and $\Gamma_{\rm n}$ generally increase with density regardless of spin constraints. When spin constraints are included, data for both simulations drop drastically, though the overall trends remain unchanged.

The spin measurement range of GW190521 is relatively wide, so incorporating spin constraints will have a smaller impact on the number of matching events in our model. In contrast, GW231123 has a high $\chi_1$, so the event rate is much lower when spin is included. As indicated in Figure \ref{fig:dmsr}, hierarchical merger BHs spin down with increasing mass; thus, as their masses approach GW231123's,  spins drop to about 0.4-0.6. This is partly attributed to our initial setup, where BH spins were initialized near zero. 
 
Furthermore, regarding the two M50k models (with and without a stellar halo potential), the presence of the halo reduces the occurrence density of IMBHs. The inclusion of the halo leads to an increase in the sum of $\Gamma_{\rm N}$ and $\Gamma_{\rm I}$ by approximately $8-927\%$ in the majority of models, while a few cases show a decrease of $26-51\%$; overall, no systematic trend is observed.

\citet{2025ApJ...994L..54P} provided $1.6 \times 10^{-3}$ - 0.16 $\text{yr}^{-1}\text{Gpc}^{-3}$ on the formation rates of events similar to GW231123 in NSCs. Using the NSC number density of $0.116 \, \text{Mpc}^{-3}$ from their work, and considering the mass-limited GW231123-like events in our M50k-IMF-ALL-r0.01 model, which occur at rates of 27.27 $\text{Gyr}^{-1}$ per NSC and 36.36 $\text{Gyr}^{-1}$ per NSC, we derive volumetric merger rates of 0.316 $\text{yr}^{-1} \, \text{Gpc}^{-3}$ and 0.422 $\text{yr}^{-1} \, \text{Gpc}^{-3}$, respectively.  We can see that the order of magnitude after conversion is consistent with the upper limit in the paper.

\begin{table}[h]
    \centering
    \caption{GW190521- and GW231123-like events produced in models.}
    \label{tab:GWs_combined}
    \sisetup{
        table-format=1.2,
        table-number-alignment=center,
        parse-numbers=false
    }
    \footnotesize
    \resizebox{1\textwidth}{!}{
    \begin{tabular}{@{} l c *{5}{c} @{\hspace{1.5em}} *{5}{c} @{}}
        \toprule
        \textbf{Model }
        & 
        & \multicolumn{5}{c}{\textbf{GW190521 ($\rbh$ [pc])}} 
        & \multicolumn{5}{c}{\textbf{GW231123 ($\rbh$ [pc])}} \\
        \cmidrule(lr){3-7} \cmidrule(lr){8-12}
         & & {0.1} & {0.05} & {0.01} & {0.005}  & {0.1} & {0.05} & {0.01} & {0.005}  \\
        \midrule
        \multirow{3}{*}{M50k-IMF-BH } 
        & $N_{\rm I}$                   & {X} & {X} & {X} & 1   
                                        & {X} & {X} & {X} & 1  \\
        & $\Gamma_{\rm I}~\rm (Gyr^{-1})$  & {X} & {X} & {X} & 0    
                                        & {X} & {X} & {X} & 0  \\
        & $N_{\rm n}$                   & 44 & 45 & 44 & 34   
                                        & 44 & 45 & 44 & 34  \\
        & $\Gamma_{\rm n}~\rm (Gyr^{-1})$  & 22.73(2.27)  & 33.33 & 84.09(4.55) & 117.65(11.76)  
                                        & 0 & 0 & 6.18  & 23.53   \\
        & $P~ (\%)$                     & 2.56(0.23)  & 2.20 & 2.73(0.17) & 3.05(0.47) 
                                        & 0 & 0 & 0.22  & 0.61  \\
        \addlinespace[0.4em]
        \cmidrule{2-12}
        \multirow{3}{*}{M50k-IMF-ALL} 
        & $N_{\rm I}$                   & {X} & {X} & 22 & 31   
                                        & {X} & {X} & 22 & 31  \\
        & $\Gamma_{\rm I}~\rm (Gyr^{-1})$  & {X} & {X} & 59.09 & 69.74(9.68)   
                                        & {X} & {X} & 27.27 & 19.35(3.23)  \\
        & $N_{\rm n}$                   & 45 & 45 & 22 & 9  
                                        & 45 & 45 & 22 & 9  \\
        & $\Gamma_{\rm n}~\rm (Gyr^{-1})$  & 11.11(4.44) & 24.44 & 90.90(13.64) & 177.78(11.11)  
                                        & 0 & 0  & 36.36  & 22.22   \\
        & $P~ (\%)$                     & 2.34(0.93) & 2.60 & 2.05(0.19) & 1.33(0.14)  
                                        & 0 & 0  & 0.87  & 0.29(0.04)   \\
        \addlinespace[0.4em]
        \bottomrule
    \end{tabular}
    }
    \vspace{0.8em}
    \begin{flushleft}
    \scriptsize
    \textbf{Note:} Values are shown as $a$ ($b$), where $a$ is from "mass-only" comparison and ($b$) is from "mass and spin" comparison.  The absence of $b$ indicates a value of zero. $N_{\rm I}$ and $N_{\rm n}$ represent the number of simulations used in our calculations. $\Gamma_{\rm I}$ and $\Gamma_{\rm n}$ denote the mean number per Gyr per cluster in IMBH-forming models and models without IMBH, respectively. $X$ denotes models in the dataset where no IMBH was formed; '0' indicates models where an IMBH exists but no events were recorded.
    \end{flushleft}
    \label{tab:GWs}
\end{table}

\subsection{The Effects of BHs on Dynamical Evolution of Clusters}

\begin{figure}[h]
    \centering
    \includegraphics[width=1\textwidth]{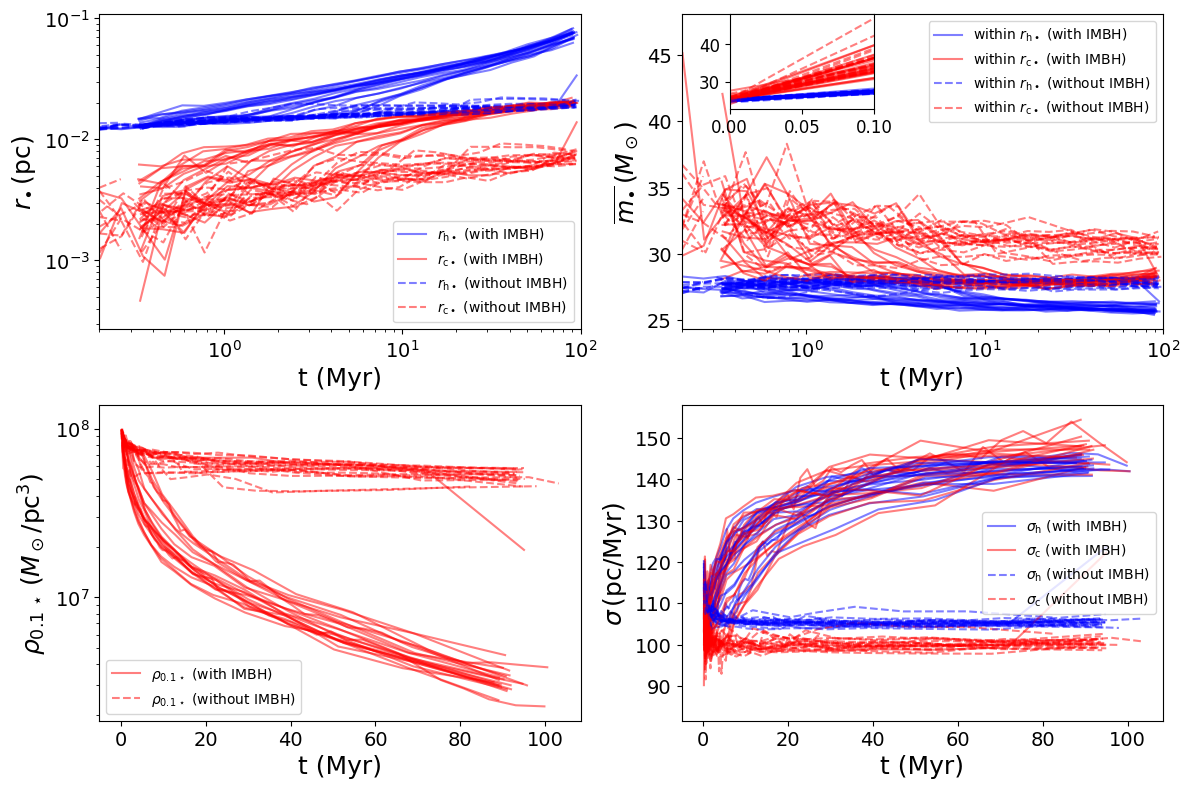}
    \caption{The half-mass radius $\rbh$ (blue), core radius $\rcbh$ (red), average mass $\mbhave$ of stellar-mass BHs ($\le 150~M_\odot$) within these radii, the stellar-mass density at 0.1 pc $\rhozstar$, the half-mass radius velocity $\sigma_{\rm h}$ (blue) and the core radius velocity dispersion $\sigma_{\rm c}$ (red) are shown as a function of time $t$ for the M50k-IMF-All-r0.01 model. Models that form an IMBH ($\ge 1000~M_\odot$) are shown with solid lines; those without an IMBH are shown with dashed lines.}
    \label{fig:cluster}
\end{figure}

The presence of an IMBH also imprints a strong signature on the global dynamics of its host cluster. Figure \ref{fig:cluster} illustrates the impact of an IMBH on the overal evolution of the star cluster. The time evolution of $\rbh$, core radius $\rc$, average mass $\mbhave$ , stellar mass density $\rho_\star$  and velocity dispersion $\sigma$, exhibits two distinct branches, depending on whether an IMBH is present. 
The existence of an IMBH causes faster cluster expansion and decreases in $\rhozstar$ within 0.1 pc and $\mbhave$ within $\rbh$ and $\rc$. This occurs because interactions between the IMBH and other BHs eject BHs to the outer regions of the cluster, thereby increasing $\rbh$. The strong potential of IMBH also increases $\sigma_{\rm c}$ and $\sigma_{\rm h}$. Furthermore, the dominance of the IMBH suppresses further low-mass BBH merger, leading to a lower $\mbhave$. Comparing $\mbhave$ at $\rbh$ and $\rc$ reveals rapid mass segregation: $\mbhave$ in the core rises sharply within 0.1 Myr, remains higher than within $\rbh$, and the ratio between them stabilizes over time.

\subsection{Comparison with observations and star formation models}
\begin{figure}[h]
    \centering
    \includegraphics[width=0.8\textwidth]{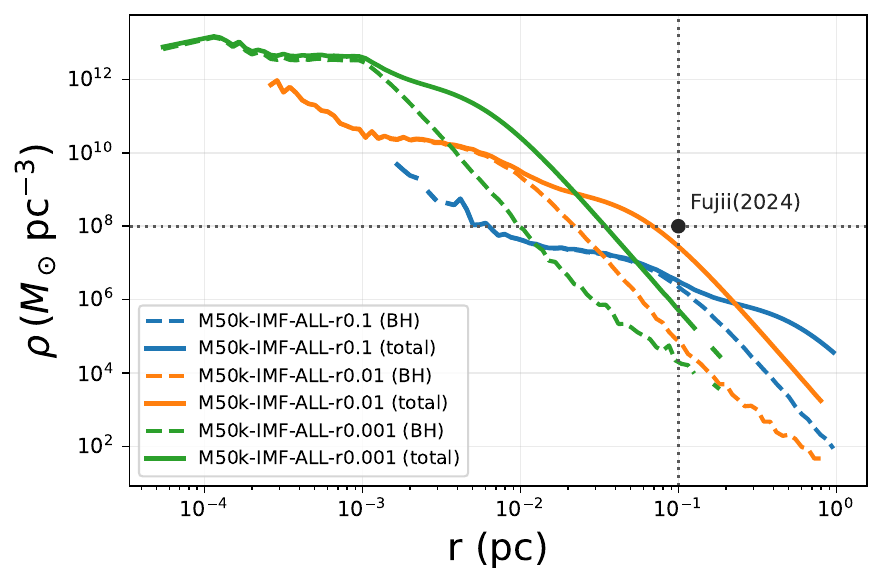}
    \caption{Density profiles as a function of radius for M50k-IMF-BH and M50k-IMF-ALL under different initial conditions with half-mass radii $\rbh$ = 0.1, 0.01, and 0.001 pc. Different colors represent different initial half-mass radii. Solid lines denote the total density profile of BHs and stellars combined for M50k-IMF-ALL, dashed lines denote the BH density profiles of M50k-IMF-ALL.}
    \label{fig:density}
\end{figure}

By comparing with observed NSCs, our model shows a reasonable stellar density after $\sim100$~Myr of evolution. As cluster evolves, $\rhozstar$ inevitably continues to decline due to ongoing IMBH growth. 
NSCs in the Milky Way and M33 have volume densities of approximately $10^7 M_\odot/\mathrm{pc}^3$ at 0.01 pc and 0.1 pc, respectively \citep{1993AJ....105.1793K,2018A&A...609A..27S}. 
In the M50k-IMF-All-r0.01 model (Figure~\ref{fig:cluster}) with IMBH formation, $\rhozstar$ decreases to $\sim 10^7 M_\odot/\mathrm{pc}^3$ at around 30 Myr, and Figure \ref{fig:generate_mass} indicates that an IMBH with $\mbh> 1000 M_\odot$ has already formed approximately 20 Myr earlier. At 0.01 pc, it takes roughly 40 Myr to reach comparable to that of observed NSCs. 
As shown in Figure 9 of \citet{2020ApJ...900...32P},  most NSCs have densities of $10^4$ to $10^6 M_\odot/\mathrm{pc}^3$ at 0.1 pc, indicating that the densities in our simulations are broadly consistent with observations. 

However, there is still a significant uncertainty remains: whether such an extremely initial high-density of BH subsystem which can form IMBHs can be acheived in nature. Figure \ref{fig:density} shows the initial density distributions as a function of radius for our different models. We adopt the Plummer model as the initial distribution for the BH subsystem. 
For models that include the stellar gravitational potential, their density distributions can be divided at the half-mass radius: the inner region is dominated by BH density, while the outer region is dominated by stellar density. In the simulations of dense GC formation \citep{2024Sci...384.1488F}, the central density reaches $10^8 M_\odot/\mathrm{pc}^3$ at 0.1 pc, which is comparable to densities in our M50k-IMF-ALL-r0.01 simulations. Thus, even if the VMS in their model does not evolve into an IMBH due to strong wind mass loss, sufficiently dense BH subsystems can still trigger runaway GW mergers to form IMBHs.

In addition, \citet{2023MNRAS.523.3201D} pointed out if the stellar free-fall timescale is shorter than the feedback delay timescale, stars can form in large numbers without being significantly affected by supernova feedback. This leads to intense star formation and produces a similarly dense population of BHs.

JWST observations also indicate that proto-GC star clusters at $z\sim10$ have densities three orders of magnitude higher than present-day young massive clusters \citep{Adamo2024,Abdurrouf2025}. 
Taken together, these results imply that such high BH densities are achievable at high redshifts.


\section{Discussion}

Our simulations do not consider gas drag effects in gas-rich systems.
\cite{2020MNRAS.498.5652K} pointed out that after stellar-mass BHs form in NSCs, gas accretion in the host galaxy and gas drag can significantly contract the BH subsystem, creating the high-density environment needed to trigger IMBH formation. Future $N$-body studies can incorporate gas effects to model this process more realistically.

Although our simulations include the stellar halo, we neglect direct stellar interactions with BHs, which can help them sink to the center via dynamical friction. We also ignore the possible ongoing supply of newly formed stellar-mass BHs from multiple star formation episodes in NSCs. Future studies should incorporate these effects in more realistic NSC simulations.

We model the general relativity effect on BBH orbital evolution using the orbit-averaged method from \cite{1963PhRv..131..435P}. Unlike post-Newtonian treatments, this approach neglects orbital precession, which is important in Kozai–Lidov triple systems; however, such triples contribute negligibly to BBH mergers in our extremely dense environments, making the approximation reasonable. We also ignore GW capture in hyperbolic encounters, which may enhance hard BBH formation and IMBH growth but is rare due to the low probability of extremely close encounters.

Previous models of IMBH formation through BH mergers all require relatively long formation timescales. Our models demonstrate that IMBH formation via BH mergers can occur within 10 Myr from the formation of the BH subsystem.
Similarly, there are other mechanisms that can also rapidly form IMBHs at high redshifts. \citet{2005ApJ...633..624V} proposed that in metal-free dark matter halos at high redshift $z\sim 20$ with virial temperature $T_{\mathrm{vir}}>10^4$ K, atomic hydrogen cooling leads to the formation of a quasi-spherical "fat" gas disk with temperature $T_{\mathrm{gas}}\sim 5000-10000$K. Seed BHs of initial mass $10^2-10^4 M_{\odot}$, formed from Pop III stellar collapse, embedded in this disk grow rapidly via supercritical Bondi accretion.
\citet{2014Sci...345.1330A} pointed out that at high redshift ($z>15$), in dense cold gas stream-fed compact star clusters (total mass $M_c=4\times 10^4 M_{\odot}$, with gas and stellar masses each accounting for half of the total mass; radius $R_c=0.25$ pc; gas pressure-supported with extremely low net angular momentum), seed black holes of initial mass $10 M_{\odot}$ grow rapidly via supra-exponential Bondi accretion in $\sim$ few $\times 10^7$ years into $\gtrsim 10^4 M_{\odot}$ BH seeds.
\citet{2021MNRAS.501.1413N} proposes that dense, gas-rich NSCs can continuously incubate and rapidly grow initial stellar-mass BHs (5--10~M$_{\odot}$) into IMBHs ($\sim 10^4$--$10^5$~M$_{\odot}$) throughout cosmic history via supra-exponential accretion and angular momentum suppression. These works all require a dense system of high-density gas with low angular momentum to ensure that black holes can maintain efficient accretion over a relatively long period. Compared with these models, our approach does not rely on substained, highly efficient gas accretion in dense, low-angular-momentum envirnoments. Instead, it requires an extremely high BH density, enabling runaway BBH mergers driven by GW radiation. If such densities can be achieved at high redshift, this provides a natural alternative channel for producing IMBH seeds and ultimately explaining the origin of high-redshift SMBHs.
Furthermore, since FFB have demonstrated that the early Universe possessed higher density environments than at present. The M50k-IMF-ALL-r0.01 model has a surface stellar density of $10^7 M_\odot /\mathrm{pc}^{2} $at 0.1 pc. If we extrapolate the Dekel model \citep{2025arXiv251107578D} inward toward the center, the density should be comparable to our model, or perhaps even higher, indicating that our model falls within the density regime achieved after wet compaction.

Several IMBH candidates (such as $\omega$ Centauri) have been identified in the local Universe, but all remain controversial: the core difficulty lies in the dynamical degeneracy. \citet{2024Natur.631..285H} discovered seven stars at the center of $\omega$ Centauri with velocities far exceeding the cluster's escape velocity, thereby strongly demonstrating the existence of an IMBH with a mass of approximately 8,200$-$47,000 $M_\odot$. 
If $\omega$ Centauri was originally the NSC of a dwarf galaxy, its central density may have satisfied the criterion ($\ge10^9 M_\odot/\text{pc}^3$) found in our simulations. Consequently, the IMBH formation scenario proposed in our work could offer a possible explanation for the IMBH observed in $\omega$ Centauri. However, this represents only one of several possible interpretations, as there is no requirement for a rapid IMBH formation in $\omega$ Centauri. Alternative mechanisms remain possible, including formation via a very massive star (VMS; \citep{2004Natur.428..724P,2024Sci...384.1488F}) or through hierarchical mergers of BBHs on a much longer timescale \citep{2015MNRAS.454.3150G,Antonini_2019}.


\section{Conclusion}

In summary, by using $N$-body simulations, this research has preliminarily validated the environmental conditions required for the runaway merger of stellar BHs to rapidlly form IMBHs. 
\begin{enumerate}
    \item The rapid early formation of an IMBH in a cluster critically depends on the total mass and initial spatial density of the BH subsystem. We observe that IMBHs can form within 10 Myr from the formation of the BH subsystem for a total cluster mass of $5\times10^4 M_\odot$, with an initial half-mass average BH density $\ge5 \times10^{9}~M_{\odot}/\text{pc}^3$ and a corresponding stellar density of $\ge10^8~M_\odot/\text{pc}^3$ at 0.1~pc. The stellar halo potential, while enhancing IMBH mass growth, introduces stochastic delays in the onset of rapid merger phases. Our simulations confirm that in BH-rich, dense star clusters, rapid dynamical formation and successive mergers of BBHs constitute a viable channel for IMBH formation. 

    \item Owing to the large velocity dispersion, three-body scattering is highly unlikely to form hard BBHs within 10~Myr. However, local fluctuations in the hard-soft boundary allow ``globally soft'' BBHs with $a$ between $10^4-10^5~ R_\odot$ to undergo temporary hardening, which can accumulate through a chain of exchanged BBHs until a final hard BBH forms at $a\sim 10^2~R_\odot$ and merges via GW radiation. 

    \item Following the formation of an IMBH, the mergers of the remaining BBHs are suppressed, making their coalescence significantly more difficult.

    \item Our models demonstrate that as IMBHs grow through hierarchical mergers, their spin $\chi$ and kick velocity $v_k$ systematically decrease. This trend is driven by the stochastic cancellation of angular momentum and the decreasing mass ratio $q \to 0$ as the primary BH outgrows its companions. In addition, BBH mergers show a positive correlation between $e$ and $a$ (Figure~\ref{fig:BHp}), driven by the GW selection effect where strong radiation at small separations efficiently circularizes the binaries during their inward spiral.

    \item We present the formation of PIBHs in our M50k-IMF-ALL-r0.001 model and a comparison between GW190521 and GW231123 based on our simulations. The formation rate of PIBHs reaches 0.7926--11.7724 per Myr per cluster within 100~Myr simulation time. As summarized in Table \ref{tab:GWs} which counts the data within 10 Myr, both $\Gamma_{\rm h}$ and $\Gamma_{\rm c}$ exhibit a positive correlation with stellar density, spanning ranges of 2.27 to 177.78 per Gyr and 3.23 to 69.74 per Gyr per cluster, respectively. These GW190521- and GW231123-like events account for approximately 0.14\%--3.05\% and 0.04\%--0.87\% of the total BBH mergers, respectively. Furthermore, the number of detectable events generally increases by 8\% to 927\% due to the stellar halo potential, though in some cases it decreases by 26\% to 51\%, with no clear overall trend.

    \item An IMBH acts as a ``dynamical engine'' that profoundly alters the evolutionary trajectory of its host cluster. Through strong dynamical interactions, it scatters stellar-mass BHs to the cluster outskirts, leading to rapid expansion of the BH subsystem (increase in half-mass radius $\rbh$) and decreases in the average BH mass $\mbhave$ within the core and the stellar-mass density $\rhozstar$ in 0.1~pc. Meanwhile, the ejection of a significant number of BHs from the subsystem core by the IMBH also leads to increases in $\sigma_{\rm h}$ and $\sigma_{\rm c}$. This feedback mechanism effectively suppresses subsequent mergers of low-mass BH binaries, shaping two distinct evolutionary branches for the cluster's structural evolution as illustrated in Figure \ref{fig:cluster}.

    \item In the typical IMBH forming model (M50k-IMF-ALL-r0.001), the stellar density decreases from $10^8$ to $10^6~M_\odot/\mathrm{pc}^3$ within 100~Myr due to IMBH heating, consistent with the observed NSC densities of Milky Way, M33 and other galaxies. The initial density matches recent simulations of dense GC formation by \citet{2024Sci...384.1488F}, and \cite{2023MNRAS.523.3201D} note the potential for even denser stellar environments. JWST observations also shows that proto-GC densities at high redshifts are far exceed those of present-day young massive clusters, supporting the plausibility of our high initial stellar density.
\end{enumerate}

This result highlights a potentially critical, yet understudied, channel for rapid IMBH formation via runaway GW mergers. Future work should validate the formation of such high-density BH subsystems using more realistic hydrodynamical simulations that incorporate gas effects and the influence of pre-existing VMS.

\begin{acknowledgments}
We thank Holger Baumgardt for providing the initial condition generator, and Renyue Cen and Douglas N. C. Lin for their valuable feedback.
Y.S. and L.W. thanks the support from the GuangDong Basic and Applied Basic Research Foundation (2026A1515012460). L.W thanks the National Natural Science Foundation of China through grant 12233013 and 12573041, the Fundamental Research Funds for the Central Universities, Sun Yat-sen University (2025QNPY04), the High-level Youth Talent Project (Provincial Financial Allocation) through the grant 2023HYSPT0706, the one-hundred-talent project of Sun Yat-sen University. The authors acknowledge the Beijing Super Cloud Center（BSCC）for providing HPC resources that have contributed to the research results reported within this paper. URL: http://www.blsc.cn/.
\end{acknowledgments}

\begin{contribution}

Y.S. implemented the GW recoil module in \textsc{petar} code, performed the simulations, analyzed the data, and wrote the manuscript. 
L.W. conceived the study, assisted with the implementation, edited the manuscript, and supervised the project.


\end{contribution}

%

\software{\textsc{petar} (\citealp{Wang_2020}, Version 1425gwkick\_307e, https://github.com/lwang-astro/PeTar),
          \textsc{sdar} (\citealp{2020MNRAS.493.3398W}, Version  1425gwkick\_307e, https://github.com/lwang-astro/SDAR),
          \textsc{fdps} (\citealp{2016PASJ...68...54I,2020PASJ...72...13I,71-Namekata.2018}, Version 7.0, https://github.com/FDPS/FDPS),
          \textsc{sse/bse} (\citealp{72-Hurley.2000,73-Hurley.2002,74-Banerjee.2020}),
          \textsc{galpy} (\citealp{2015ApJS..216...29B}, modified version: https://github.com/lwang-astro/galpy),
          TwoPlummer (private communication with Holger Baumgardt), 
          Precession \citep{2016PhRvD..93l4066G},
          numpy (\citealp{85-Harris.2020}),
          matplotlib (\citealp{86-Hunter.2007}),
          astropy (\citealp{astropy2013..2013A&A...558A..33A,astropy2018..2018AJ....156..123A,astropy2022..2022ApJ...935..167A})
          }


\appendix

\section{Implementing the GW Recoil Model}
\label{sec:GWkick}

In GW-driven BBH inspirals, the coupling between BH spins and orbital angular momentum introduces long-term precession: both spins and the orbital plane precess around the system's total angular momentum \citep{Apostolatos1994, PhysRevD.52.821}. As energy and momentum are emitted as GWs, the orbital separation gradually shrinks, ultimately leading to a merger \citep{1963PhRv..131..435P}. Due to asymmetries in component masses and spins, the remnant BH receives a recoil kick, typically on the order of $10^2-10^3$ km~$\text{s}^{-1}$. The \textsc{percession} code models BBH inspirals  using orbit- and precession-averaged approaches \citep{Gerosa_2016} and employs fitting formulas derived from numerical relativity simulations to predict the remnant's mass, spin, and recoil velocity.

We begin by introducing the basic framework of the formulas. All quantities are expressed in units where $c = G = m =1$, with total BBH mass $\Mbh = m_1 + m_2$; mass ratio $q = m_2/m_1~ (m_2 \le m_1)$; symmetric mass ratio $\eta = m_1m_2/\Mbh^2 = q/(1+q)^2$; and component spin magnitudes $S_i = m^2_i\chi~(i = 1,2)$, where $\chi_i$ are dimensionless spin parameters ranging from 0 to 1. The fitting formulas are generally expressed in terms of the following weighted combinations of the BH spins:

\begin{align}
            \bm{\Delta}=\frac{q \chi_2\hat{\mathbf{S_2}} -\chi_1\hat{\mathbf{S_1}}}{1+q}, \quad \tilde{\bm{\chi}} =  \frac{q^2\chi_2\hat{\mathbf{S_2}}+\chi_1\hat{\mathbf{S_1}}}{(1+q)^2} \label{eq:basic1} 
\end{align}

The recoil arises from asymmetries in the masses or spins of the merging BHs. The mass asymmetry contribution, $v_{\mathrm{m}}$, lies within the orbital plane, while the spin-induced recoil has components parallel and perpendicular to the orbital angular momentum, denoted as $v_{\parallel}$ and $v_{\perp}$, respectively. The total recoil velocity magnitude, $\vk$, can be modeled as \citep{Campanelli_2007}:

\begin{align}
    \vk = \sqrt{v_\mathrm{m}^2+2v_\mathrm{m}v_{\perp}cos\zeta+v_{\perp}^2 +v^2_{\parallel}} \label{eq:vkick}    
\end{align}
where $\zeta$ is the angle between the mass term and the in-plane spin term. The expressions for $v_\mathrm{m} ,v_{\perp}$ and $v_{\parallel}$ are given by:

\begin{align}
    v_{\mathrm m} &= A\eta^2 \frac{1-q}{1+q} (1+B\eta) \label{eq:vm} \\
    v_{\parallel} &=H\eta^2 \Delta_{\parallel}   \label{eq:vparallel} \\
    v_{\perp} &= 16\eta^2[\Delta_{\perp}(V_{11}+2V_A \tilde{\chi_{\parallel}}+4V_B\tilde{\chi_{\parallel}}^2+8V_C \tilde{\chi_{\parallel}}^3)+2 \tilde{\chi_{\perp}}\Delta_{\parallel}(C_2+2C_3\tilde{\chi_{\parallel}})]cos\Xi \label{eq:vperp}
\end{align}
where $\Delta_{\parallel}$, $\Delta_{\perp}$, $\tilde{\chi_{\parallel}}$, $\tilde{\chi_{\perp}}$ represent the components of $\bm{\Delta}$ and $\tilde{\bm{\chi}}$ parallel and perpendicular to the spin-orbit angular momentum $\bm{L}$. The phase angle $\Xi$ is a random value between 
0 and $2\pi$, depending on the initial relative positions of the BBH components. The corresponding parameter values are: $A = 1.2\times 10^4 ~\text{km~s}^{-1}$, $B=-0.93$ \citep{PhysRevLett.98.091101}, $H = 6.9\times10^3~\text{km~s}^{-1}$ \citep{PhysRevD.77.044028}, $V_{11}=3677.76~\text{km~s}^{-1}$, $V_\mathrm{A} = 2481.21~\text{km~s}^{-1}$, $V_\mathrm{B}=1792.45~\text{km~s}^{-1}$, $V_\mathrm{C}=1506.52~\text{km~s}^{-1}$ \citep{Lousto2012GravitationalRF}, $C_2 =1140~\text{km~s}^{-1}$, $C_3=2482~\text{km~s}^{-1}$ \citep{PhysRevD.87.084027}, $\zeta=145^{\circ}$ \citep{PhysRevD.77.044028}.

We implement these key fitting formulas \citep[see details in ][]{Gerosa_2016} as a module in \textsc{petar}, working together with \textsc{bse} module. When a BBH meets the GW merger criterion defined by \textsc{bse}, the module calculates the remnant BH's mass, velocity and three-dimensional spin. Since \textsc{bse} records only the spin magnitude, the initial spin direction of a newly formed BH is assigned by randomly sampling from a uniform distribution.

\bibliography{references}
\bibliographystyle{aasjournalv7}


\end{CJK*}
\end{document}